\documentclass[prl,superscriptaddress,twocolumn]{revtex4}
\usepackage{latexsym}
\usepackage{amsfonts}
\usepackage{amssymb}
\usepackage{amsmath}
\usepackage{graphicx}

\begin{document}

\title{Vertical current induced domain wall motion in MgO-based magnetic tunnel junction with low current densities}

\author{A. Chanthbouala}
\affiliation{Unit\'e Mixte de Physique CNRS/Thales and Universit\'e Paris Sud 11, 1 ave A. Fresnel, 91767 Palaiseau, France}
\author{R. Matsumoto}
\affiliation{Unit\'e Mixte de Physique CNRS/Thales and Universit\'e Paris Sud 11, 1 ave A. Fresnel, 91767 Palaiseau, France}
\author{J. Grollier}
\email{julie.grollier@thalesgroup.com}
\affiliation{Unit\'e Mixte de Physique CNRS/Thales and Universit\'e Paris Sud 11, 1 ave A. Fresnel, 91767 Palaiseau, France}
\author{V. Cros}
\affiliation{Unit\'e Mixte de Physique CNRS/Thales and Universit\'e Paris Sud 11, 1 ave A. Fresnel, 91767 Palaiseau, France}
\author{A. Anane}
\affiliation{Unit\'e Mixte de Physique CNRS/Thales and Universit\'e Paris Sud 11, 1 ave A. Fresnel, 91767 Palaiseau, France}
\author{A. Fert}
\affiliation{Unit\'e Mixte de Physique CNRS/Thales and Universit\'e Paris Sud 11, 1 ave A. Fresnel, 91767 Palaiseau, France}
\author{A. V. Khvalkovskiy}
\altaffiliation{Now at Grandis, Inc., 1123 Cadillac Court, Milpitas, CA 95035, USA}
\affiliation{Unit\'e Mixte de Physique CNRS/Thales and Universit\'e Paris Sud 11, 1 ave A. Fresnel, 91767 Palaiseau, France}
\affiliation{A.M. Prokhorov General Physics Institute of RAS, Vavilova str. 38, 119991 Moscow, Russia}
\author{K.A. Zvezdin}
\affiliation{A.M. Prokhorov General Physics Institute of RAS, Vavilova str. 38, 119991 Moscow, Russia}
\affiliation{Istituto P.M. s.r.l., via Cernaia 24, 10122 Torino, Italy}
\author{K. Nishimura}
\affiliation{Process Development Center, Canon ANELVA Corporation, Kurigi 2-5-1, Asao, Kawasaki, Kanagawa 215-8550, Japan}
\author{Y. Nagamine}
\affiliation{Process Development Center, Canon ANELVA Corporation, Kurigi 2-5-1, Asao, Kawasaki, Kanagawa 215-8550, Japan}
\author{H. Maehara}
\affiliation{Process Development Center, Canon ANELVA Corporation, Kurigi 2-5-1, Asao, Kawasaki, Kanagawa 215-8550, Japan}
\author{K. Tsunekawa}
\affiliation{Process Development Center, Canon ANELVA Corporation, Kurigi 2-5-1, Asao, Kawasaki, Kanagawa 215-8550, Japan}
\author{A. Fukushima}
\affiliation{National Institute of Advanced Industrial Science and Technology (AIST) 1-1-1 Umezono, Tsukuba, Ibaraki 305-8568, Japan}
\author{S. Yuasa}
\affiliation{National Institute of Advanced Industrial Science and Technology (AIST) 1-1-1 Umezono, Tsukuba, Ibaraki 305-8568, Japan}

\maketitle

\textbf{Shifting electrically a magnetic domain wall (DW) by the spin transfer mechanism \cite{Slonczewski:JMMM:1996,Berger:PRB:1996,Grollier:APL:2003,Klaui:APL:2003} is one of the future ways foreseen for the switching of spintronic memories or registers \cite{Parkin:Science:2008,NEC}. The classical geometries where the current is injected in the plane of the magnetic layers suffer from a poor efficiency of the intrinsic torques \cite{Hayashi:PRL:2007,Klaui:PRL:2005} acting on the DWs. A way to circumvent this problem is to use vertical current injection \cite{Ravelosona:PRL:2006,Boone:PRL:2010,Rebei:PRB:2006}. In that case, theoretical calculations \cite{Khvalkovskiy:PRL:2009} attribute the microscopic origin of DW displacements to the out-of-plane (field-like) spin transfer torque \cite{Slonczewski:PRB:2005,Theodonis:PRL:2006}. Here we report experiments in which we controllably displace a DW in the planar electrode of a magnetic tunnel junction by vertical current injection. Our measurements confirm the major role of the out-of-plane spin torque for DW motion, and allow to quantify this term precisely. The involved current densities are about 100 times smaller than the one commonly observed with in-plane currents \cite{Lou:APL:2008}. Step by step resistance switching of the magnetic tunnel junction opens a new way for the realization of spintronic memristive devices \cite{Strukov:Nature:2008,Wang:IEEE:2009,Grollier:patent:2010}. }

We devise an optimized sample geometry for efficient current DW motion using a magnetic tunnel junction with an MgO barrier sandwiched between two ferromagnetic layers, one free, the other fixed. Such junctions are already the building block of magnetic random-access memories (M-RAMs), which makes our device suitable for memory applications. The large tunnel magnetoresistance \cite{Yuasa:NatMat:2004,Parkin:NatMat:2004} allows us to detect clearly DW motions when they propagate in the free layer of the stack \cite{Kondou:APEX:2008}. The additional advantage of  magnetic tunnel junctions is that the out-of-plane field-like torque $\mathbf{T_{OOP}}$ can reach large amplitudes, up to 30$\%$ of the classical in-plane torque $\mathbf{T_{IP}}$ \cite{Sankey:Nature:2007,Kubota:Nature:2007}, in contrast to metallic spin-valve structures, in which the out-of-plane torque is only a few $\%$ of the in-plane torque \cite{Stiles:PRB:2002,Xia:PRB:2002}. This is of fundamental importance since theoretical calculations predict that, when the free and reference layers are based on materials with the same magnetization orientation (either in-plane or perpendicular), the driving torque for steady domain wall motion by vertical current injection is the OOP field-like torque \cite{Khvalkovskiy:PRL:2009}. Indeed, $\mathbf{T_{OOP}}$ is equivalent to the torque of a magnetic field in the direction of the reference layer, that has the proper symmetry to push the DW along the free layer. On the contrary, the in-plane torque $\mathbf{T_{IP}}$ can only induce a small shift of the DW of a few nm. In magnetic tunnel junctions with the same composition for the top free and bottom reference layers, the OOP field-like torque exhibits a quadratic dependence with bias \cite{Sankey:Nature:2007,Kubota:Nature:2007}, which could not allow us to reverse the DW motion by current inversion. Therefore we use asymmetric layer composition to obtain an asymmetric OOP field-like torque \cite{Oh:Nature:2009,Tang:PRB:2010}.

\begin{figure*}
	\includegraphics[width=.7\textwidth]{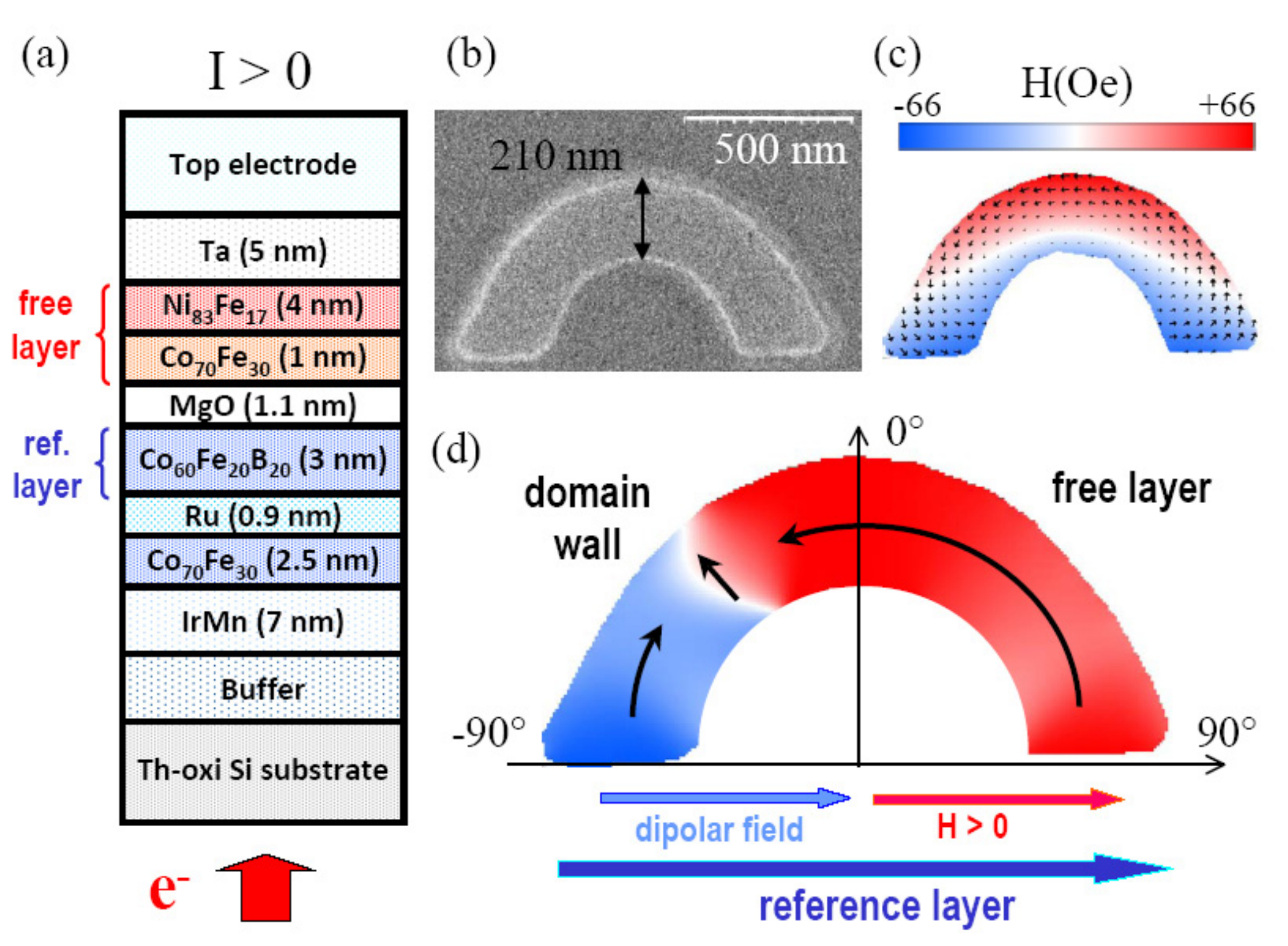}
	\caption{Magnetic tunnel junction design for DW motion by vertical current injection (a) Sketch of the MgO-based magnetic tunnel junction stack.  (b)  S.E.M. image corresponding to a top view of the junction before adding the top electrode. The width of this sample is 210 nm. (c) Micromagnetic simulations giving the distribution of the Oersted field induced by the perpendicular current for a current density of 4 10$^6$ A.cm$^{-2}$. The color scale corresponds to the amplitude of the magnetization projected on the long axis. (d) Schematic diagram of the sample geometry and micromagnetic simulations showing the stable DW position.}
	\label{fig1}
\end{figure*}	

The magnetic stack is sketched in Fig.\ref{fig1} (a).   The top free layer is (CoFe 1nm/NiFe 4 nm), and the fixed layer is a CoFeB alloy. An S.E.M. top view image of the sample geometry before adding the top contact is shown on Fig.\ref{fig1} (b). The half-ring shape was designed for two reasons. First, it facilitates the DW creation \cite{Saitoh:Nature:2004}. As can be seen from the micromagnetic simulations presented on Fig.\ref{fig1} (d), the larger width at the edges stabilizes the DW at an intermediate position in the wire. Secondly, it allows a specific distribution of the Oersted field created by the perpendicular current, as shown by the simulations of Fig.\ref{fig1} (c). Thanks to the hollow center, the Oersted field is quasi unidirectional along the wire, and can assist the DW propagation.  

\begin{figure*}
	\includegraphics[width=.7\textwidth]{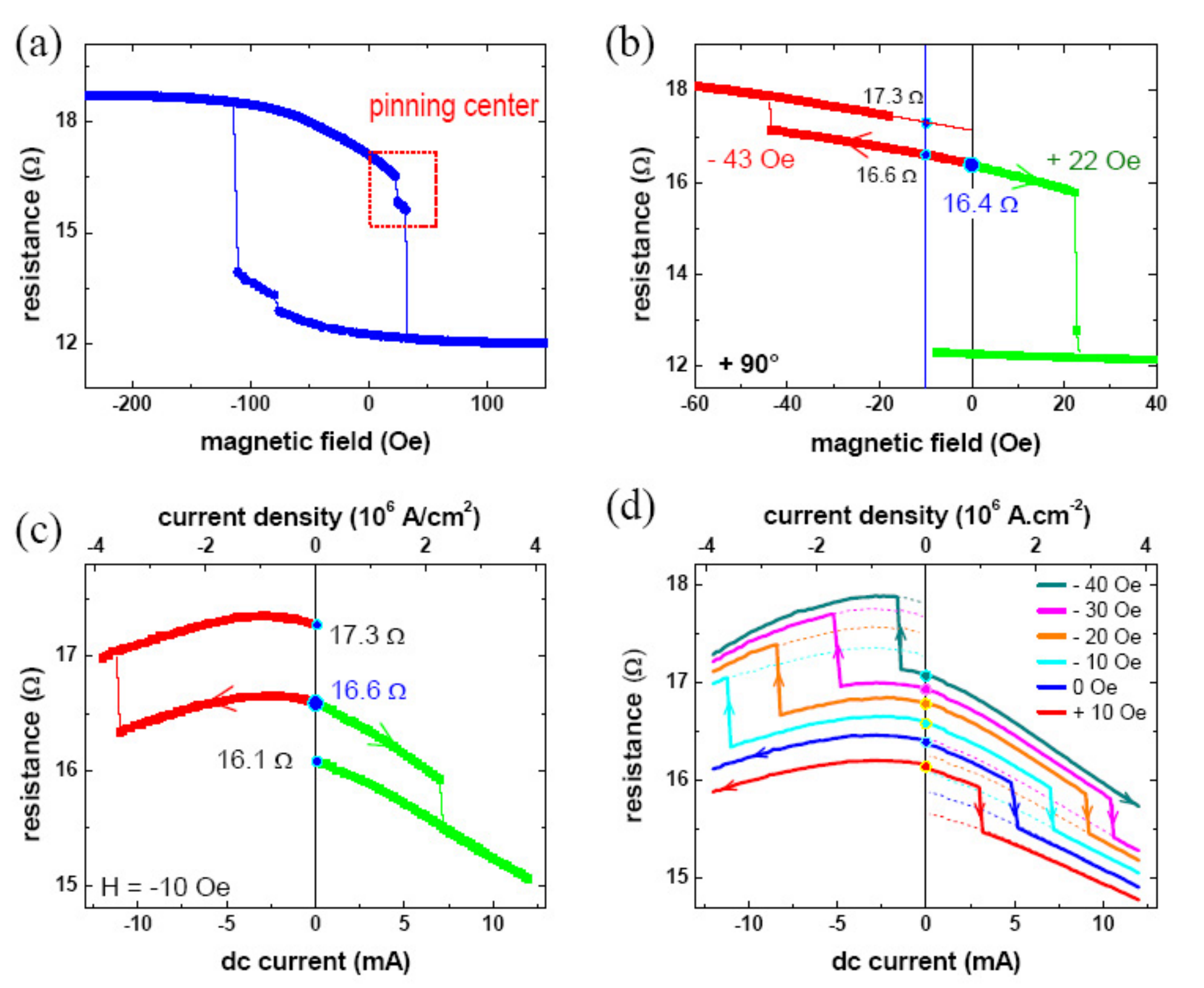}
	\caption{Vertical current-induced DW depinning (a) Resistance versus magnetic field curves measured with the field applied along 75$^{\circ}$. (b) Resistance versus field curves obtained with the DW initially pinned at zero field (R = 16.4 $\Omega$). The field is applied along 90$^{\circ}$. The green (resp. red) curve is obtained by applying positive (resp. negative) fields. The depinning fields are $H_{dep}^+$ = +22 Oe and $H_{dep}^-$ = - 43 Oe. (c) Resistance versus current curves obtained with the DW initially pinned. The applied field is - 10 Oe, the initial state for the two curves is R = 16.6 $\Omega$. The green (resp. red) curve is obtained by applying positive (resp. negative) currents. (d) Resistance versus current curves obtained with the DW initially pinned. Each curve is measured with a fixed applied magnetic field between - 40 and + 10 Oe. The curves for positive and negative currents are obtained independently, the initial DW state is reset between each curve. In (c) and (d), the bottom axis gives the applied dc current, while the top axis corresponds to the injected current density.}
	\label{fig2}
\end{figure*}

We first focus on the results obtained with the 210 nm wide wires. A sketch of the sample geometry is given in Fig.\ref{fig1} (d), including our convention for the angle of the applied magnetic field. In order to create and pin a DW, we tilt the magnetic field to 75$^{\circ}$. As can be seen in in Fig.\ref{fig2} (a), plateaus appear in the resistance vs. field R(H) curve, corresponding to the creation of a magnetic domain wall close to the sample edge (as in the micromagnetic simulation of Fig.\ref{fig1} (d)). We chose to work with the plateau obtained at positive fields ($\approx$ + 15 Oe) close to the AP state, which is stable when the field is swept back to zero. This DW creation/pinning process is reproducible, allowing measurements with the same initial state. The strength of the pinning can be evaluated by measuring the corresponding depinning fields. After pinning the DW and coming back to zero field, the R(H) curves have been measured by increasing the field amplitude along 90$^{\circ}$, either to negative or positive values, as shown in Fig.\ref{fig2} (b). The positive (resp. negative) depinning fields are $H_{dep}^+$ = +22 Oe and $H_{dep}^-$ = - 43 Oe. This indicates an asymmetry of the potential well which is due to the dipolar field of the synthetic antiferromagnet ($\approx$ + 40 Oe) and also to the asymmetric geometry of the sample close to the edge.

In order to study the current induced domain wall depinning, once the domain wall is created, we apply a fixed magnetic field between $H_{dep}^+$ and $H_{dep}^-$, for example - 10 Oe, corresponding to zero effective field, as illustrated by a blue vertical line in Fig.\ref{fig2} (b). In our convention, a positive current corresponds to electrons flowing from the synthetic antiferromagnet to the free layer. In Fig.\ref{fig2} (c), we show two resistance versus current curves obtained at - 10 Oe, starting always from the same initial DW position (resistance 16.6 $\Omega$). In addition to the expected decrease of the tunnel resistance with bias, we clearly observe irreversible resistance jumps. When the current is swept first to positive values (green curve), the resistance is switched at $I_{dep}^+$ = + 7 mA to a lower resistance state corresponding to another domain wall position, stable at zero current, with a low bias resistance of 16.1 $\Omega$. By resetting the DW position, then applying negative currents (red curve) a resistance jump to a higher resistance state of 17.3 $\Omega$ occurs at $I_{dep}^-$ = -11 mA. We thus demonstrate the possibility to move a domain wall by perpendicular dc current injection in both directions depending on the current sign. The current densities corresponding to the DW motion are lower than 4 10$^6$ A.cm$^{-2}$ (see top x axis of Fig.\ref{fig2} (c)). The use of perpendicular current injection therefore allows to reduce the current densities by a factor 100 compared to the classical lateral current injection \cite{Hayashi:PRL:2007,Klaui:PRL:2005}. 

\begin{figure*}
	\includegraphics[width=.7\textwidth]{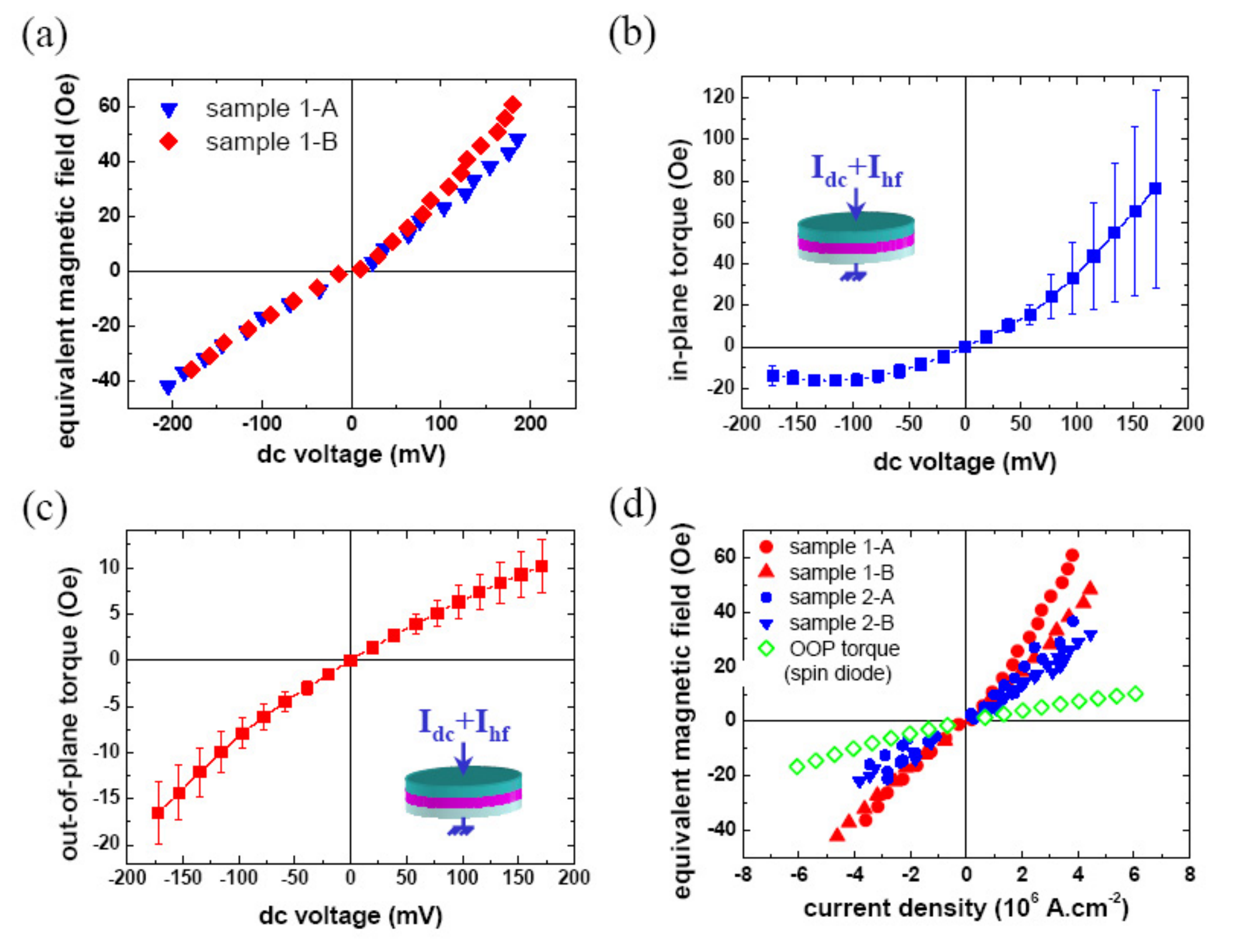}
	\caption{ Spin torques measurements : DW depinning versus spin diode (a) Plot of the equivalent field versus dc voltage obtained by the DW depinning experiments, for two similar samples of width 210 nm (sample 1-A, blue symbols, sample 1-B red). The curves (b) and (c) are obtained from spin diode experiments performed with 270 $\times$ 70 nm$^2$ elliptic samples etched in the same stack. (b) In-plane torque and (c) out-of-plane field-like torque as a function of dc voltage. (d) Plot of the equivalent field versus current density obtained by the DW depinning experiments, for the two samples of type 1 (width 210 nm, inner diameter 550 nm, filled red symbols), and the two smaller samples of type 2 (width 120 nm, inner diameter 370 nm, filled blue symbols). For comparison, the OOP field-like torque obtained from spin diode experiments is represented in open green symbols.}
	\label{fig3}
\end{figure*}	

Similar measurements have been performed for several fields between $H_{dep}^+$ and $H_{dep}^-$. As shown on Fig.\ref{fig2} (d), the resistance associated with each pinning center changes progressively as a function of the applied magnetic field, which can be ascribed to field-induced DW displacement / deformation within the potential well. The depinning currents strongly depend on the applied magnetic field too. Negative fields favour the domain wall motion in the -90$^{\circ}$ direction, thus reducing the values of $I_{dep}^-$, and increasing $I_{dep}^+$. As expected, the effect is opposite for positive fields. By comparing with the DW motion in applied field, we can define the value of the equivalent field induced by the positive or negative depinning currents : $I_{dep}^{\pm}$ in field $H$ generates an equivalent field of $H_{dep}^{\pm}$-$H$. We therefore can plot the equivalent field generated by the current as a function of the bias voltage, as shown in Fig.\ref{fig3} (a) for two samples with the same nominal shape. Additional experiments allow us to discard Joule heating (which could reduce the current-induced depinning fields) as a possible source of measured effective field enhancement at large bias (see methods). For both samples, a positive bias induces an effective field pointing in the direction of the reference layer magnetization and inversely for negative bias. The overall trend is similar, linear at low bias ($<$ 60 mV), with deviations from linearity at large bias. 

The origins of current-induced DW depinnings are : the two spin-transfer torques (in-plane or out-of-plane) and the Oersted field. Spin diode experiments are a powerful tool to obtain the dc bias dependence of the two spin transfer torques \cite{Sankey:Nature:2007,Kubota:Nature:2007,Wang:PRB:2009}. In order to investigate the physical origin of the perpendicular current-induced domain wall motion in our system, we perform additional spin diode experiments with 70 $\times$ 270 nm$^2$ ellipses patterned in the same stack as the semi-circular wires. The analysis of the resulting rectified voltage vs frequency curves obtained at different bias allow us to plot the two components of the spin transfer torque, the in-plane torque Fig.\ref{fig3} (b) and the out-of-plane field-like torque Fig.\ref{fig3} (c), expressed in field units, as a function of bias (see methods and supplementary information). As expected, the IP torque is asymmetric with bias. The sign of the OOP torque of our asymmetric structure also changes with the current direction, in agreement with the results of Oh \textit{et al.} obtained by another method in MgO-based tunnel junctions with asymmetric layer compositions. In the low bias region between $\pm$ 60 mV, the OOP field-like torque reaches up to 40 $\%$ of the IP torque. 

\begin{figure}
	\includegraphics[width=.45\textwidth]{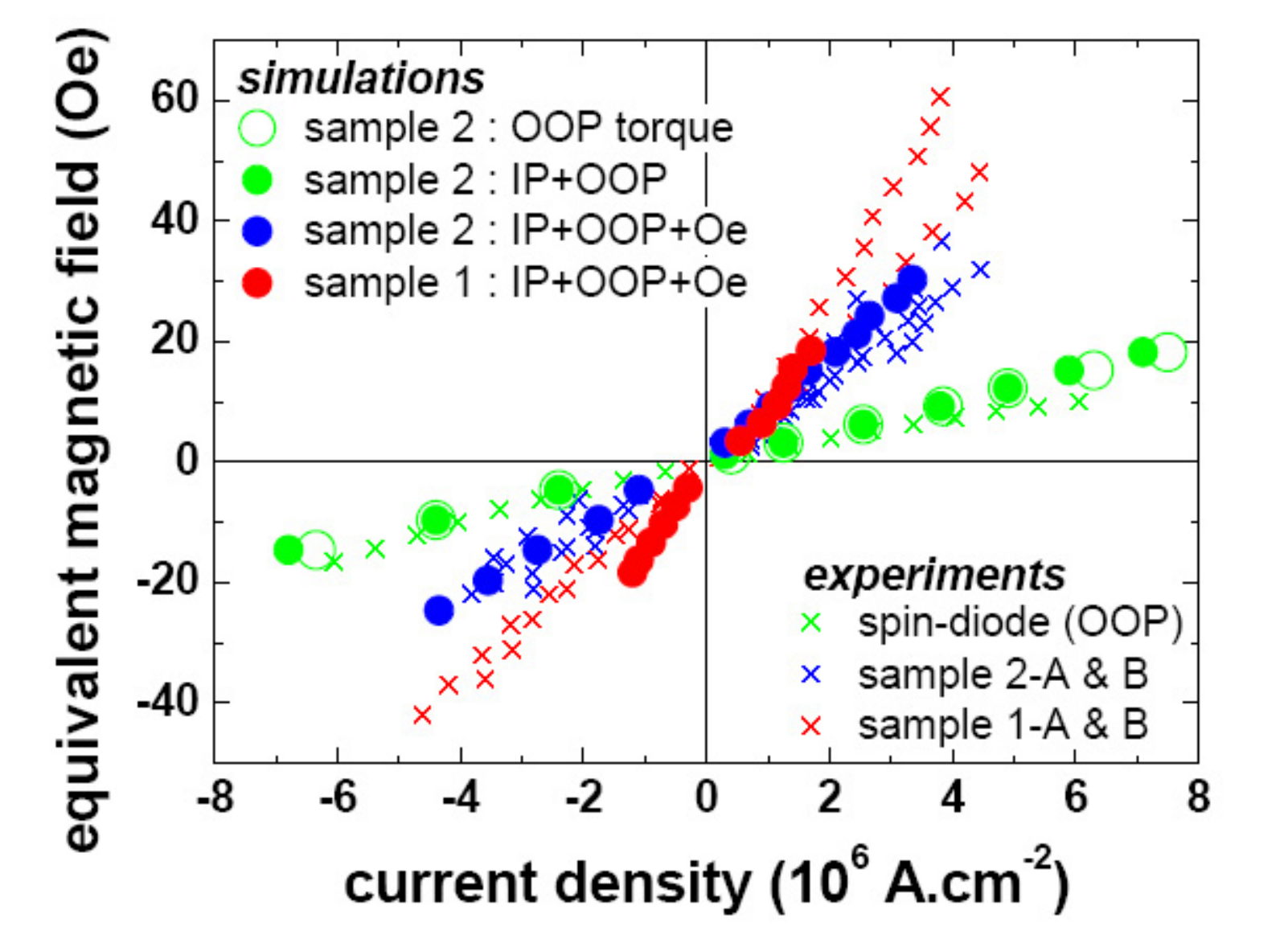}
	\caption{Plot of the equivalent field versus dc voltage. Circle symbols : micromagnetic simulations of DW depinning. Red and blue cross symbols :  experimental results for the current-induced depinning fields. Green crosses : OOP torque expressed in field units derived from spin diode experiments.}
	\label{fig4}
\end{figure}	

According to our previous theoretical predictions \cite{Khvalkovskiy:PRL:2009}, only the OOP term can give rise to spin transfer induced steady DW motion. Fig.\ref{fig3} (d) directly compares the amplitude of the OOP torque (expressed in field units) derived from spin diode experiments to the effective field determined by domain wall depinning measurements. For both types of experiments, a positive bias induces an equivalent field pointing in the direction of the magnetization of the reference layer, and in the opposite direction for negative bias. Samples 1-A and B have the dimensions corresponding to the S.E.M. picture of Fig.\ref{fig1} (b) (width 210 nm), while samples 2-A and B are smaller but with the same aspect ratio (width 120 nm). For the largest samples 1-A and 1-B, the equivalent magnetic field generated by perpendicular injection is clearly increased at high bias. 

In order to interpret these results, we have performed additional micromagnetic simulations following the experimental procedure of domain wall depinning (supplementary note 3). The spin transfer torques and the Oersted field are introduced in order to determine the equivalent current induced magnetic fields. In the simulations, both spin transfer torques are set linear as a function of current, with $T_{OOP} = 40 \% \:T_{IP}$. The results are summed up on Fig.\ref{fig4}. First ignoring the Oersted field, we obtain approximately the same equivalent fields with only $T_{OOP}$ (open green circles) than with both $T_{IP}$ and $T_{OOP}$ (filled green circles), which confirms our predictions \textit{et al.} \cite{Khvalkovskiy:PRL:2009} that the contribution of $T_{IP}$ to the DW motion is negligible. These simulations are in very good agreement with the OOP torque derived from the spin diode experimental data (green crosses). We find that the contribution of the spin transfer torques to the equivalent field (closed green circles) is not sufficient to account for the experimental equivalent fields for samples 1 and 2 (blue and red crosses), but that a quantitative agreement can be obtained by taking the Oersted field also into account (blue and red filled circles). In particular, this allows us to ascribe unambiguously the larger slope obtained for samples 1-A and B to the increased Oersted field for larger areas. These results also prove the efficiency of our approach combining the actions of the Oersted field and OOP torque to induce DW propagation at current densities lower than 5 10$^6$ A.cm$^{-2}$, thanks to the specific design of our sample. This torque engineering could be of particular interest for the development of the low current density multilevel memory cells proposed by Seagate \cite{Lou:APL:2008}.

From Fig.\ref{fig3} (d) and Fig.\ref{fig4}, it appears that the OOP field-like torque alone can generate an equivalent magnetic field of 10 Oe for 5 10$^6$ A.cm$^{-2}$. Devices using only this mechanism to drive DW motion are therefore possible if the DW is not strongly pinned. This last condition is typically desirable for a DW based spintronic memristor, in which the DW position should be continuously tunable by current injection \cite{Wang:IEEE:2009,Grollier:patent:2010}. Memristor devices inherently behave like artificial nano-synapses and have a strong potential for implementation in large-scale neuromophic circuits \cite{Strukov:Nature:2008}. The more intense is the current through a memristor, and the longer it is injected, the more the resistance changes. Spin-transfer induced DW displacements are precisely proportional to the amplitude of the injected current as well as pulse duration \cite{Hayashi:PRL:2007}. In addition, by using perpendicularly magnetized layers (domain wall width $<$ 10 nm \cite{Ravelosona:PRL:2006}), our device could be scaled down below 50x100 nm$^2$. Low current density DW motion by perpendicular current injection in large TMR magnetic tunnel junctions is therefore extremely promising for the future developments of fast and robust spintronic memristors.

\textbf{Methods} :

SAMPLES

The magnetic stack was grown by sputtering in a Canon ANELVA chamber. Details of the growth and fabrication process have been presented elsewhere \cite{Yuasa:JPhysD:2007}. For all samples the TMR is around 65 $\%$, with a low RA product of 3.5 $\Omega.\mu m^2$. Samples 1-A and B have the dimensions corresponding to the S.E.M. picture of Fig.\ref{fig1} (b) (width 210 nm, inner diameter 550 nm), while samples 2-A and B are smaller but with the same aspect ratio (width 120 nm, inner diameter 370 nm).

TEMPERATURE EVALUATION

In order to evaluate the increase of temperature in our samples, we have measured the saturation fields $ H_{SAT} $ of the synthetic antiferromagnet at constant low bias as a function of the temperature, and as a function of bias at constant temperature (RT). By comparing the two sets of measurements, we estimate that the largest temperature increase is $\approx$ 20 K and has a negligible impact on the DW depinning fields ($\approx$ 1 Oe) (see supplementary information).

SPIN-TORQUE DIODE MEASUREMENTS

The magnetic field is applied along the hard axis of the ellipse, and is chosen large enough to saturate the magnetization of the free layer (experimental applied field range 550 - 650 0e). We inject a constant microwave current of $i_{hf}$ = 40 $\mu A$ and sweep the dc current between -0.9 and +0.9 mA. The microwave current is modulated (on/off 1:1) in order to increase the precision. 

MICROMAGNETIC SIMULATIONS

For the micromagnetic simulations, we use the finite-difference micromagnetic code SpinPM, developed by Istituto P.M. The simulated free layer has the geometry of the S.E.M. images of samples 1 or 2. The mesh cell size is set to 3 $\times$ 3 $\times$ 5 nm$^{3}$. We took the following magnetic parameters: $\alpha$ = 0.01 for the Gilbert damping and $M_{s}(CoFe/NiFe)$ = 1 T for the magnetization of the free layer. The spin polarization has been set to P$_{spin}$ = 0.5 and the current is supposed to be uniform through the structure.

\textbf{Acknowledgments} : 

Financial support by the CNRS, RFBR grant (Grant No. 09-02-01423), JSPS Postdoctoral Fellowships for Research Abroad and the European Research Council (Starting Independent Researcher Grant No. ERC 2010 Stg 259068) is acknowledged.
Correspondence and requests for materials should be addressed to J.G.

\textbf{Author Contributions} : 

J.G., A.C., V.C. and S.Y. conceived the experiments. A.C. and R.M. carried out the measurements and analyzed the data with the help of J.G. and V.C.; A.C. performed the numerical simulations with help from J.G., A.V.K. and K.A.Z.; K.N., Y.N., H.M, K.T. deposited the magnetic stack. A.F. fabricated the samples. J.G. wrote the paper with discussions and comments from A.C., R.M., V.C., A.A., S.Y. and A.F.

\end{document}